\documentclass[multphys,vecphys]{svmult}

\usepackage{url}             
\usepackage{makeidx}         
\usepackage{graphicx}        
\usepackage{multicol}        
\usepackage{cite}            
\usepackage[bottom]{footmisc}

\makeindex             

\begin{document}

\title{Theory of the Quantum Hall Effect in Quasi-One-Dimensional
Conductors}
\titlerunning{Theory of the Quantum Hall Effect}
\author{Victor M.\ Yakovenko}
\institute{Department of Physics, University of Maryland, College
Park, MD 20742-4111, USA, \url{http://www2.physics.umd.edu/~yakovenk}}

\maketitle

This Chapter reviews the theory of the quantum Hall effect (QHE) in
quasi-one-dimensional (Q1D) conductors.  It is primarily based on the
author's papers \cite{Yakovenko91,Yakovenko96,Yakovenko98,Sengupta01}.
The QHE in Q1D conductors is closely related to the
magnetic-field-induced spin-density wave (FISDW) observed in these
materials.  The theory of the FISDW is reviewed in this book by
L.~Gor'kov, by M.~H\'eritier, by A.~Lebed, and by S. Haddad et al.
The FISDW experiment is reviewed by P.~Chaikin et al.\ and by
V.~Pudalov and A.~Kornilov.

\section{Introduction to quasi-one-dimensional conductors}
\label{Sec:Intro}

Organic metals of the (TMTSF)$_2$X family, where TMTSF is
tetramethyltetraselenafulvalene, and X is an inorganic anion such as
PF$_6$, are Q1D crystals consisting of parallel conducting chains
formed by the organic molecules TMTSF.  The chains direction is
denoted as \vec{a} or $x$.  The interchain coupling is much weaker in
the \vec{c} ($z$) direction than in the \vec{b} ($y$) direction, so
the chains form weakly coupled two-dimensional (2D) layers.  In a
simple model, the electron dispersion $\varepsilon(\vec{k})$ in these
materials is described by a tight-binding model representing tunneling
between the TMTSF molecules
\begin{equation}
  \varepsilon = 2t_a\cos(k_xa)+2t_b\cos(k_yb)+2t_a\cos(k_zd)+\ldots\ts,
\label{dispersion}
\end{equation}
where $\vec{k}=(k_x,k_y,k_c)$ is the electron wave vector.  The
intermolecule distances are $a=0.73$ nm, $b=0.77$ nm, and $d=1.35$ nm
\cite{Ishiguro-book}, and we approximate the triclinic crystal
structure by the orthogonal one.  The electron tunneling amplitudes
$t_a \gg t_b \gg t_c$ are estimated as 250 meV, 25 meV, and 1.5 meV
\cite{Ishiguro-book}.  The band (\ref{dispersion}) is
quarter-filled\footnote{We ignore weak dimerization of the TMTSF
molecules \cite{Ishiguro-book}, which is not essential for our
consideration.} by holes, because each anion X$^-$ takes one electron.
The Fermi surface is open and consists of two disconnected sheets with
$k_x$ close to $\pm k_\mathrm{F}$, where $k_\mathrm{F}=\pi/4a$ is the
Fermi momentum along the chains.  In the vicinity of the Fermi
surface, we can linearize the longitudinal electron dispersion
$\varepsilon_\|(k_x)$.  Measuring $\varepsilon$ from the Fermi energy
and neglecting $t_c$, we can approximate (\ref{dispersion}) as
\begin{equation}
  \varepsilon_\pm = \varepsilon_\|(k_x)+\varepsilon_\perp(k_yb) =
  \pm\hbar v_\mathrm{F}(k_x\mp k_\mathrm{F}) + 2t_b\cos(k_yb) 
  + 2t_b'\cos(2k_yb) \ts,
\label{linearized}
\end{equation}
where the signs $\pm$ correspond to the two sheets of the Fermi
surface, $v_\mathrm{F}=\partial\varepsilon_\|/\hbar\partial
k_x\approx10^5$~m/s \cite{Takahashi05} is the Fermi velocity, and
$\hbar$ is the Planck constant.  In this Chapter, we study the
in-plane Hall conductivity $\sigma_{xy}$ per one layer in a magnetic
field \vec{B} applied perpendicular to the \vec{a}--\vec{b} layers.
The interlayer coupling $t_c$ is not essential for this consideration,
so most of the theory is presented for just one 2D layer, as in
(\ref{linearized}).  The tunneling amplitude $t_b'$ to the
next-nearest chain in (\ref{linearized}) is important for the FISDW
theory.

\section{Hall effect in the normal state}
\label{Sec:Normal}

Let us briefly discuss the Hall effect in the normal state of
(TMTSF)$_2$X.  The textbook formula says that the Hall coefficient is
$R_\mathrm{H}=1/nec$, where $n$ is the carrier concentration per one
layer, $e$ is the electron charge, and $c$ is the speed of light.
However, the linearized model (\ref{linearized}) gives
$R_\mathrm{H}=0$, because of the electron-hole symmetry.  A non-zero
result is obtained by taking into account the curvature $\beta$ of the
longitudinal electron dispersion
\cite{Cooper,Virosztek89,Zheleznyak99}:
\begin{equation}
  R_\mathrm{H}^{\rm(n)}=\frac{\beta}{nec} \ts, \quad
  \beta=\frac{k_\mathrm{F}}{v_\mathrm{F}
}\,
  \frac{\partial^2\varepsilon_\|}{\partial k_x^2} \ts.
\label{beta}
\end{equation}
For the quarter-filled tight-binding band (\ref{dispersion}) in
(TMTSF)$_2$X, $n=1/2ab$ and $\beta=\pi/4$
\cite{Cooper,Virosztek89,Zheleznyak99}, as opposed to $\beta=1$ for
the conventional parabolic dispersion.  The experimentally measured
$R_\mathrm{H}^{\rm(n)}$ \cite{Forro00,Moser00} is in overall agreement
with (\ref{beta}), but it also exhibits some puzzling temperature
dependence \cite{Zheleznyak99,Zheleznyak01}.

The Hall resistivity
$\rho_{xy}=R_\mathrm{H}^{\rm(n)}B=\beta(h/e^2)(2abB/\phi_0)$ is small
and not quantized.  This is because, for any realistic $B$, the
magnetic flux through the area $2ab$ per one carrier is much smaller
that the flux quantum $\phi_0=hc/e$, so the Landau filling factor is
very high, of the order of $10^2$--$10^3$.

\section{Introduction to the quantum Hall effect in the 
  FISDW state}
\label{Sec:FISDW+QHE}

When a strong magnetic field $B$ is applied in the $z$ direction
perpendicular to the layers, the (TMTSF)$_2$X materials experience a
cascade of phase transitions between the so-called
magnetic-field-induced spin-density-wave states.  These are true
thermodynamic phase transitions, observed in specific heat
\cite{Pesty85,Fortune90}, magnetization \cite{Naughton85}, NMR
\cite{Takahashi84}, and virtually all other physical quantities
\cite{Scheven95,McKernan95}.  A detailed theory of the FISDW is
presented in other chapters of this book, as well as in the book
\cite{Ishiguro-book} and in the review volume \cite{Schegolev}.
According to the theory
\cite{Heritier84,Montambaux85,Gorkov85,Lebed85}, the electron spin
density in the FISDW state develops spontaneous modulation with the
wave vector
\begin{equation}
  \vec{Q}=(Q_x,Q_y,Q_z)\ts, \quad Q_x=2k_\mathrm{F}-NG\ts, \quad
  G=ebB/\hbar c, \quad E_B=\hbar v_\mathrm{F}G \ts,
\label{Q}
\end{equation}
where $G$ is the characteristic wave vector of the magnetic field, and
$N$ is an integer number (positive or negative).  The magnetic length
$l_x=2\pi/G$ is defined so that the magnetic flux through the area
bounded by $l_x$ and by the interchain distance $b$ is equal to the
flux quantum: $Bl_xb=\phi_0$.  Notice that $G\ll k_\mathrm{F}$,
because $Bab\ll\phi_0$ for realistic magnetic fields.  We also
introduced the characteristic energy $E_B$ of the magnetic field in
(\ref{Q}).

The difference between $Q_x$ and $2k_\mathrm{F}$ in (\ref{Q}) plays
crucial role for the Hall effect.  Let us calculate $\sigma_{xy}$ in
the FISDW state by naively counting the number of available carriers
\cite{Heritier84}.  The FISDW with the wave vector \vec{Q}
(\ref{Q}) hybridizes the $\pm k_\mathrm{F}$ sheets of the Fermi
surface and opens an energy gap.  Some electron and hole pockets form
above and below the energy gap because of imperfect nesting.  The
effective carrier concentration $n$ is determined by their total area
in the momentum space.  This is the area between the two sheets of the
Fermi surface (\ref{linearized}) $k_x^{\pm}(k_y)=\pm k_\mathrm{F} \mp
\varepsilon_\perp(k_y)/v_\mathrm{F} $, one of them shifted by the wave
vector $(Q_x,Q_y)$:
\begin{eqnarray}
  n &=& \frac{2}{2\pi}\int_0^{2\pi/b}\frac{\D k_y}{2\pi}\{
  k_x^+(k_y)-[k_x^-(k_y-Q_y)+Q_x]\} 
\nonumber \\
  &=& \frac{2}{2\pi}\int_0^{2\pi/b}\frac{\D k_y}{2\pi}
  \left( 2k_\mathrm{F}-Q_x + 
   \frac{\varepsilon_\perp(k_y-Q_y)-\varepsilon_\perp(k_y)}
   {v_\mathrm{F}}
  \right) \ts.
\label{pockets}
\end{eqnarray}
Here the factor 2 accounts for two spin projections.  The integral of
the last term in (\ref{pockets}) vanishes for any $Q_y$, because
\begin{equation}
  \int_{0}^{2\pi/b}\varepsilon_\perp(k_y)\,\D k_y=0 \ts,
\label{E_perp}
\end{equation}
whereas (\ref{Q}) and the first term in (\ref{pockets}) give
$n=2NeB/hc$.  Substituting this expression in the textbook formula
$\sigma_{xy}=nec/B$, we find
\begin{equation}
  \sigma_{xy}=2Ne^2/h \ts.
\label{s_xy}
\end{equation}
Formula (\ref{s_xy}) was also derived in \cite{Poilblanc87} using the
St\u{r}eda formula.

Equation (\ref{s_xy}) represents the integer quantum Hall effect
(QHE).  The Hall effect becomes quantized, because the FISDW (\ref{Q})
eliminates most of the carriers, but leaves $N$ fully occupied Landau
levels in the remaining pockets \cite{Chaikin92}.  With the increase
of $B$, the system experiences a cascade of phase transitions between
the FISDW states with different numbers $N$.  This produces a series
of the quantum Hall plateaus (\ref{s_xy}) separated by phase
transitions, as indeed observed experimentally in (TMTSF)$_2$X
\cite{Chaikin92,Hannahs88,Jerome88,Valfells96,Chamberlin88,Naughton88,Kang91,Jerome91b,Balicas95,Cho99}.
In a simple case, $N$ takes consecutive numbers 0, 1, 2, \ldots\ with
decreasing magnetic field $B$
\cite{Chaikin92,Hannahs88,Jerome88,Valfells96}, but $N(B)$ may also
show more complicated sequences, including sign changes
\cite{Jerome88,Chamberlin88,Naughton88,Kang91,Jerome91b,Balicas95,Cho99}.
The bulk Hall conductivity is measured in experiment, and
$\sigma_{xy}$ per one layer is obtained by dividing by the number of
conducting layers.  The latter may depend on current distribution, so
the absolute values of $\sigma_{xy}$ are not obtained very precisely
\cite{Kang91}, but the relative values of $\sigma_{xy}$ do correspond
to integer numbers.  In the FISDW state,
$\sigma_{xy}\gg\sigma_{xx},\:\sigma_{yy}$, as expected for the QHE
\cite{Jerome88,Chamberlin88,Naughton88,Balicas95,Cho99}.

The derivation of (\ref{s_xy}) presented above is simple, but not
quite satisfactory.  In the FISDW state, there are multiple gaps in
the energy spectrum of electrons, including a gap at the Fermi level,
so the notion of metallic electron and hole pockets is not quite
meaningful.  Moreover, the magnetic energy $E_B$ is typically greater
than the FISDW gap $\Delta$, so there is strong magnetic breakdown
between the pockets, and the semiclassical Landau quantization using
closed electron orbits is not well defined.  After briefly summarizing
the FISDW theory in Sect.\,\ref{Sec:FISDW}, we present a more rigorous
derivation of the QHE as a topological invariant \cite{Yakovenko91} in
Sect.\,\ref{Sec:QHE}.  This formalism is also utilized for various
generalizations presented in the subsequent Sections.

\section{Mathematical theory of the FISDW}
\label{Sec:FISDW}

A magnetic field $B$ applied perpendicular to the layers can be
introduced in the electron dispersion (\ref{linearized}) via the
Peierls--Onsager substitution $k_y\rightarrow k_y-(e/\hbar c)A_y$ in
the gauge $A_y=Bx$.  As a result, the transverse dispersion
$\varepsilon_\perp(k_yb-Gx)$ produces a periodic potential in the $x$
direction with the wave vector $G$ given by (\ref{Q})
\cite{Gorkov84,Chaikin85}.  Electrons also experience the periodic
potential $\Delta_0\cos(Q_xx+Q_yy)$ from the FISDW.  To find the
energy spectrum, let us decompose the electron wave function
$\psi(x,k_y)$ into the components $\psi_+(x,k_y)$ and $\psi_-(x,k_y)$
with the longitudinal momenta close to $\pm k_\mathrm{F}$:
\begin{equation}
  \psi(x,k_y)=\psi_+(x,k_y)\,\E^{+\I k_\mathrm{F}x} 
  + \psi_-(x,k_y)\,\E^{-\I k_\mathrm{F}x}\ts.
\label{+-k_F}
\end{equation}
Introducing the two-component spinor
$[\psi_+(x,k_y),\psi_-(x,k_y+Q_y)]$, we can write the electron
Hamiltonian as a $2\times2$ matrix operating on this spinor
\begin{equation}
  \hat{H}= \left( \begin{array}{cc}
  -\I v_\mathrm{F}
\partial_x + \varepsilon_\perp(k_yb-Gx) 
  & \Delta_0\E^{\I(Q_x-2k_\mathrm{F})x}\\
  \Delta_0\E^{-\I(Q_x-2k_\mathrm{F})x} & \I v_\mathrm{F}
\partial_x + 
  \varepsilon_\perp(k_yb+bQ_y-Gx)  \end{array} \right) \ts.
\label{H1}
\end{equation}
Here the diagonal terms represent the electron dispersion
(\ref{linearized}) in the presence of a magnetic field.  The
off-diagonal terms represent the periodic potential of the FISDW
(\ref{Q}).  The wave vector $2k_\mathrm{F}$ is subtracted from $Q_x$
because of the $\E^{\pm \I k_\mathrm{F}x}$ factors introduced in
(\ref{+-k_F}).  We do not write the spin indices explicitly in
(\ref{+-k_F}) and (\ref{H1}), because they are not essential for our
consideration.  Equations (\ref{+-k_F}) and (\ref{H1}) can be applied
to pairing between $\psi_+$ and $\psi_-$ with parallel or antiparallel
spins, as discussed in more detail in \cite{Yakovenko91}.

Now let us make a phase transformation of the spinor
components\footnote{This kind of transformation was first introduced
in \cite{Gorkov84}, which started development of the FISDW theory.}
\begin{equation}
  \left( \begin{array}{l} \psi_+ \\ \psi_- \end{array} \right) =
  \left( \begin{array}{l} \psi'_+ \,
  \exp\left[(\I/E_B)\int^{k_yb-Gx}\varepsilon_\perp(\xi)\,\D\xi\right]\\
  \psi'_- \,
  \exp\left[-(\I/E_B)\int^{k_yb-Gx}\varepsilon_\perp(Q_yb+\xi)\,
  \D\xi\right]
  \end{array} \right) \ts,
\label{pe}
\end{equation}
where $E_B$ is the characteristic magnetic energy given in (\ref{Q}).
Substituting (\ref{pe}) into (\ref{H1}), we obtain a new Hamiltonian
acting on the spinor $(\psi'_+,\psi'_-)$
\begin{equation}
  \hat{H}'= \left( \begin{array}{ll}
  -\I v_\mathrm{F}
\partial_x   & \tilde\Delta(x) \\
  \tilde\Delta^*(x) & \I v_\mathrm{F}
\partial_x \end{array} \right) \ts.
\label{H'}
\end{equation}
As a result of the transformation (\ref{pe}), the $\varepsilon_\perp$
terms are removed from the diagonal in (\ref{H'}), but they re-appear
in the off-diagonal terms
\begin{equation}
  \tilde\Delta(x)=\Delta_0 \exp\left\{-\I NGx 
  - \frac{\I}{E_B}\int^{k_yb-Gx}
  [\varepsilon_\perp(\xi)+\varepsilon_\perp(Q_yb+\xi)]\,\D\xi
  \right\} \ts.
\label{D(x)}
\end{equation}
Since $\varepsilon_\perp(k_y)$ satisfies (\ref{E_perp}), the integral
in (\ref{D(x)}) is a periodic function of $k_yb-Gx$, so
$\tilde\Delta(x)$ can be expanded in a Fourier series with
coefficients $c_m$
\begin{equation}
  \tilde\Delta(x)=\Delta_0\,\E^{-\I NGx}\sum_m c_m\,\E^{\I m(k_yb-Gx)}\ts.
\label{Fourier}
\end{equation}
Because the FISDW forms at a low transition temperature in high
magnetic fields, it is appropriate to consider the limit $\Delta_0\ll
E_B$, where the FISDW potential is much weaker than the magnetic
energy.  In this case, when (\ref{Fourier}) is substituted into
(\ref{H'}), each periodic potential in the sum (\ref{Fourier}) can be
treated separately and opens a gap $2|\Delta_0c_m|$ in the energy
spectrum at the wave vectors $k_x=\pm[k_\mathrm{F}-(N+m)G/2]$
\cite{Poilblanc86,Virosztek86}.  The electron mini-bands separated by
the energy gaps can be interpreted as the broadened Landau levels,
with the number of states in each mini-band proportional to $G\propto
B$.

The term with $m=-N$ in (\ref{Fourier}) opens a gap at the Fermi
level.  This is possible only if $2k_\mathrm{F}-Q_x$ is an integer
multiple of $G$, i.e.\ there are $N$ mini-bands between the gaps at
$k_x=\pm Q_x/2$ and $k_x=\pm k_\mathrm{F}$.  Let us focus on the gap
at the Fermi level and omit the terms with $m\neq-N$ in the sum
(\ref{Fourier}).  In this single-gap approximation
\cite{Poilblanc86,Virosztek86}, the Hamiltonian (\ref{H'}) becomes the
same as for a 1D density wave \cite{Gruner94} with the effective
amplitude $\Delta \E^{-\I\varphi}$:
\begin{equation}
  \hat{H}'= \left( \begin{array}{ll} -\I v_\mathrm{F}
\partial_x &
  \Delta\,\E^{-\I\varphi(k_y)} \\ 
  \Delta^*\E^{\I\varphi(k_y)} & \I v_\mathrm{F}
\partial_x
  \end{array} \right) \ts, \quad \Delta=\Delta_0c_{-N} \ts, \quad
  \varphi=Nbk_y \ts.
\label{1-gap}
\end{equation}
The Hamiltonian (\ref{1-gap}) has the gapped energy spectrum
\begin{equation}
  E(p_x)=\sqrt{(v_\mathrm{F}
p_x)^2+|\Delta|^2} \ts, 
  \quad p_x=\hbar(k_x-k_\mathrm{F}) \ts.
\label{E(p_x)}
\end{equation}
Notice that the off-diagonal terms in (\ref{1-gap}) have the phase
$\varphi(k_y)$, which does not matter for the energy spectrum
(\ref{E(p_x)}), but plays crucial role in the QHE.\footnote{The
electron conductivity tensor for the FISDW state was calculated in
\cite{Chang87}, but quantized contribution to $\sigma_{xy}$ was lost.}

The Fourier coefficients $c_m$ (\ref{Fourier}) depend on $Q_y$ and the
ratio of the tunneling amplitudes $t_b$ and $t_b'$ (\ref{linearized})
to the magnetic energy $E_B$.  The values of $N$ and $Q_y$ in a FISDW
state are selected in such a way as to maximize the coefficient
$|c_{-N}(Q_y)|$ and, thus, the energy gap $\Delta$ (\ref{1-gap}) at
the Fermi level for a given magnetic magnetic field.\footnote{The
single-gap approximation is not sufficient for a self-consistent
calculation of thermodynamic quantities, such as the free energy and
magnetization \cite{Montambaux88}.  However, it is adequate for
describing low-energy electron states relevant for the QHE.}  As
magnetic field changes, the optimal values of $N$ and $Q_y$ change.
Since the parameter $N$ must be integer in order to produce a gap at
the Fermi level, it changes by discontinuous jumps, which produces a
cascade of the FISDW transitions.

\section{Quantum Hall effect as a topological invariant}
\label{Sec:QHE}

Suppose an electric field $E_y$ is applied perpendicular to the
chains.  We can introduce it in the Hamiltonian (\ref{H1}) by the
Peierls--Onsager substitution $k_y\rightarrow k_y-(e/\hbar c)A_y$
using the gauge $A_y=-E_yct$, where $t$ is time.  Then, the periodic
potential $\varepsilon_\perp[k_yb-G(x-vt)]$ starts to move with the
velocity $v=cE_y/B$.  This motion induces some current $j_x$ along the
chains, which constitutes the Hall effect.  However, the FISDW
periodic potential in the off-diagonal terms in (\ref{H1}) does not
move, if the FISDW is pinned.  The moving and non-moving periodic
potentials are combined in the effective Hamiltonian (\ref{1-gap}),
where the phase becomes time-dependent $\varphi=Nb(k_y+eE_yt/\hbar)$.
The time-dependent phase means that the effective 1D density wave
(\ref{1-gap}) slides along the chains, carrying the Fr\"ohlich current
\cite{Gruner94}
\begin{equation}
  j_x=\frac{2e}{2\pi b}\,\frac{\partial\varphi}{\partial t}
  =\frac{2Ne^2}{h}\,E_y=\sigma_{xy}E_y \ts.
\label{j_x}
\end{equation}
Equation (\ref{j_x}) represents the quantum Hall effect and agrees
with (\ref{s_xy}).\footnote{A more rigorous treatment of two periodic
potentials is presented in \cite{Yakovenko96,Yakovenko98}, using the
methods of \cite{Zak85,Kunz86,Kohmoto93} and giving the same result
(\ref{j_x}).}

The Hall conductivity at zero temperature can be also expressed in
terms of a topological invariant called the Chern number
\cite{Thouless,Kohmoto89}:
$$
  \sigma_{xy}=-\I\frac{e^2}{\hbar}\sum_a
  \int\frac{\D k_x}{2\pi}\int\frac{\D k_y}{2\pi}
  \left( \frac{\partial\langle\psi_a|}{\partial k_x}
  \frac{\partial|\psi_a\rangle}{\partial k_y}-
  \frac{\partial\langle\psi_a|}{\partial k_y}
  \frac{\partial|\psi_a\rangle}{\partial k_x} \right)
$$
\begin{equation}
  =-\I\frac{e^2}{\hbar}\sum_a
  \int\frac{\D k_x}{2\pi}\int\frac{\D k_y}{2\pi}
  \left[ \frac{\partial}{\partial k_x} \left(\langle\psi_a|
  \frac{\partial|\psi_a\rangle}{\partial k_y}\right) -
  \frac{\partial}{\partial k_y} \left(\langle\psi_a|
  \frac{\partial|\psi_a\rangle}{\partial k_x}\right) \right] \ts.
\label{curl}
\end{equation}
Here $|\psi_a(k_x,k_y)\rangle$ are the normalized eigenvectors of the
Hamiltonian.  The integral is taken over the Brillouin zone, and the
sum over $a$ goes over all completely occupied bands, assuming there
are no partially filled bands.

Let us apply (\ref{curl}) to the FISDW state \cite{Yakovenko91}, first
setting $Q_y=0$.  The eigenfunctions of (\ref{1-gap}) are defined on
the Brillouin zone torus $-k_\mathrm{F}\leq k_x\leq k_\mathrm{F}$ and
$0\leq k_y \leq 2\pi/b$ with the gap $\Delta\,\E^{-\I\varphi(k_y)}$ at
$k_x=\pm k_\mathrm{F}$.  Let us start with a wave function
$|\psi_0\rangle$ at some point $k_x^0$ far away from $\pm
k_\mathrm{F}$ and change $k_x$ along a closed line encircling the
torus at a fixed $k_y$.  As we pass through $+k_\mathrm{F}$, the wave
function transforms from $\psi_+$ to $\E^{\I\varphi}\psi_-$.  The
phase factor appears because the off-diagonal terms in (\ref{1-gap})
have the phase.  When we return to the original point
$k_x^0+2k_\mathrm{F}$ in the next Brillouin zone, the wave function
becomes $\E^{\I\varphi}|\psi_0\rangle$.  The first term in
(\ref{curl}) is a full derivative in $k_x$, so it reduces to a
difference taken between $k_x^0+2k_\mathrm{F}$ and $k_x^0$
$$
  -\I\frac{e^2}{h}\int_0^{2\pi/b} \frac{\D k_y}{2\pi}\left(
  \langle\psi_0|\E^{-\I\varphi(k_y)}
  \frac{\partial}{\partial k_y}\,
  \E^{\I\varphi(k_y)}|\psi_0\rangle
  -\langle\psi_0|\frac{\partial}{\partial k_y}\,|\psi_0\rangle\right)
$$
\begin{equation}
  =\frac{e^2}{h}\int_0^{2\pi/b} \frac{\D k_y}{2\pi}
  \frac{\partial \varphi(k_y)}{\partial k_y}\,
  =[\varphi(2\pi/b)-\varphi(0)]\,\frac{e^2}{h}=\frac{Ne^2}{h} \ts.
\label{phase}
\end{equation}
Multiplied by the spin factor 2, (\ref{phase}) gives the same result
as (\ref{s_xy}).

The second term in (\ref{curl}) gives zero.  In this term, the
expression under the integral can be rewritten as the difference
$\langle\psi|\partial_{k_x}|\psi\rangle|^{k_y=2\pi/b}_{k_y=0}$.  The
Hamiltonian (\ref{1-gap}) at $k_yb=2\pi$ is the same as at $k_y=0$,
and we can select the wave functions to be the same, thus the
difference equals zero.  It was shown in \cite{Yakovenko91} that the
results do not change when we take into account $Q_y\neq0$,
$Q_z\neq0$, and the multiple gaps below the Fermi energy generated by
the periodic potentials in (\ref{Fourier}).

\section{Coexistence of several order parameters}
\label{Sec:manyFISDW}

The formalism presented in Sect.\,\ref{Sec:QHE} is particularly useful
in the case where several FISDWs with different amplitudes $\Delta_j$
and numbers $N_j$ coexist \cite{Yakovenko91}.  In this case, the
off-diagonal terms in (\ref{1-gap}) become
\begin{equation}
  \Delta(k_y)=\sum_j \Delta_j \exp(-ibk_yN_j) \ts.
\label{Dky}
\end{equation}
According to (\ref{phase}), the Hall conductivity is determined by the
winding number of the complex function (\ref{Dky}), i.e.\ by the
number of times the phase of $\Delta(k_y)$ changes by $2\pi$ when
$k_y$ goes from 0 to $2\pi/b$.  This integer number, taken with the
opposite sign, must be substituted in (\ref{s_xy}) instead of $N$.
So, when several FISDWs coexist, $\sigma_{xy}$ is not a superposition
of partial Hall conductivities, but is always given by the integer
winding number of (\ref{Dky}).

When two FISDWs coexist, $\sigma_{xy}$ is given by the integer $N_j$
whose partial gap $|\Delta_j|$ is bigger.  To illustrate this, let us
use a vector representation of complex numbers and a planetary
analogy.  Let us associate the first term in the sum (\ref{Dky}) with
a vector pointing from the Sun to the Earth, and the second term with
a vector from the Earth to the Moon.  As the parameter $k_y$
increases, the Earth rotates around the Sun, and the Moon rotates
around the Earth.  The Hall conductivity is determined by the number
of times the Moon rotates around the Sun.  Clearly, this winding
number is determined only by the bigger orbit of the Earth.  If one
partial gap $|\Delta_l|$ is bigger then the sum of all other partial
gaps $|\Delta_l|>\sum_{j\neq l} |\Delta_j|$, then $\sigma_{xy}$ is
determined only by the biggest term in (\ref{Dky}), i.e.\ $N=N_l$ in
(\ref{s_xy}).

A FISDW state consisting of multiple periodic potentials was discussed
in \cite{Machida90}, and the QHE in this model was studied in
\cite{Yakovenko94a}.  Lebed pointed out in \cite{Lebed90} that the
umklapp scattering requires coexistence of two FISDWs with $N$ and
$-N$.  The QHE in this case was studied in \cite{Dupuis98a} using the
topological method, and it was found that $\sigma_{xy}$ may take the
values corresponding to $N$, $-N$, or zero.  A more detailed study was
presented in \cite{Dupuis98b}.  It was suggested that this effect was
may explain sign reversals of the QHE observed in $\rm(TMTSF)_2PF_6$
\cite{Jerome88,Balicas95,Cho99}.  An alternative theory of the QHE
sign reversals was proposed in \cite{Montambaux96}.  Sign changes of
the QHE are also observed in $\rm(TMTSF)_2ClO_4$
\cite{Chamberlin88,Naughton88,Kang91} and $\rm(TMTSF)_2ReO_4$
\cite{Jerome91b}, which have anion ordering.  Coexistence of the FISDW
pairings between different branches of the folded Fermi surface in
this case was proposed in \cite{McKernan95,Scheven95}.

\section{Temperature evolution of the quantum Hall effect}
\label{Sec:temperature}

Equation (\ref{j_x}) is a good starting point for discussion of the
temperature dependence of the QHE
\cite{Yakovenko96,Yakovenko98,Yakovenko96b}.  According to this
equation, the Hall conductivity can be viewed as the Fr\"ohlich
conductivity of the effective 1D density wave (\ref{1-gap}).  Thus,
the temperature dependence of the QHE must be the same as the
temperature dependence of the Fr\"ohlich conductivity, which was
studied in the theory of density waves \cite{Lee79,Maki90}.  At
$T\neq0$, the electric current carried by the density-wave condensate
is reduced with respect to the zero-temperature value by a factor
$f(T)$, which also reduces $\sigma_{xy}$:
\begin{equation}
  \sigma_{xy}(T)=f(T)\,2Ne^2/h \ts,
\label{f(T)2Ne2/h}
\end{equation}
\begin{equation}
  f(T)=1-\int_{-\infty}^\infty \frac{\D p_x}{v_\mathrm{F}}
  \left(\frac{\partial E}{\partial p_x}\right)^2
  \left[-\frac{\partial n_{\rm F}(E/T)}{\partial E}\right] \ts,
\label{f(T)}
\end{equation}
where $E(p_x)$ is the electron dispersion (\ref{E(p_x)}) in the FISDW
state, and $n_\mathrm{F}(\epsilon/T)=(\E^{\epsilon/T}+1)^{-1}$ is the
Fermi distribution function with $k_\mathrm{B}=1$.  Equations
(\ref{f(T)2Ne2/h}) and (\ref{f(T)}) have a two-fluid interpretation.
The first term in (\ref{f(T)}) represents the FISDW condensate current
responsible for the QHE.  The second term represents the normal
component originating from electron quasiparticles thermally excited
above the energy gap.  They equilibrate with the immobile crystal
lattice and do not participate in the Fr\"ohlich current, thus
reducing the Hall coefficient.  A simple derivation of (\ref{f(T)}) is
given in
\cite{Yakovenko96,Yakovenko98,Yakovenko96b}.\footnote{Reference
\cite{Virosztek89} discussed the Hall conductivity in the FISDW state
at $T\neq0$, but failed to reproduce the QHE at $T=0$.}

\begin{figure}
\hfill \includegraphics*[width=0.35\textwidth,angle=-90]{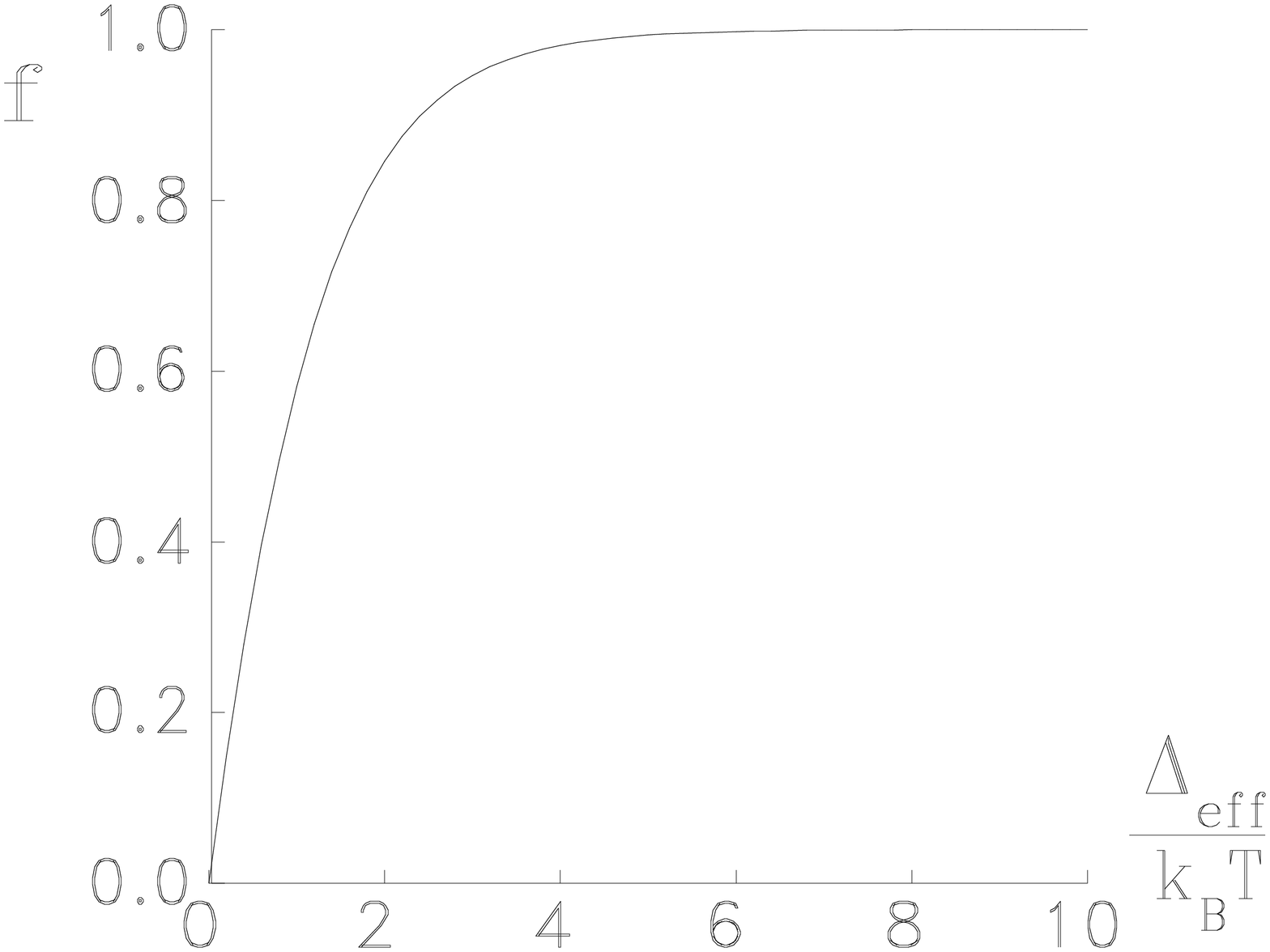}
\hfill\hfill
\includegraphics*[width=0.35\textwidth,angle=-90]{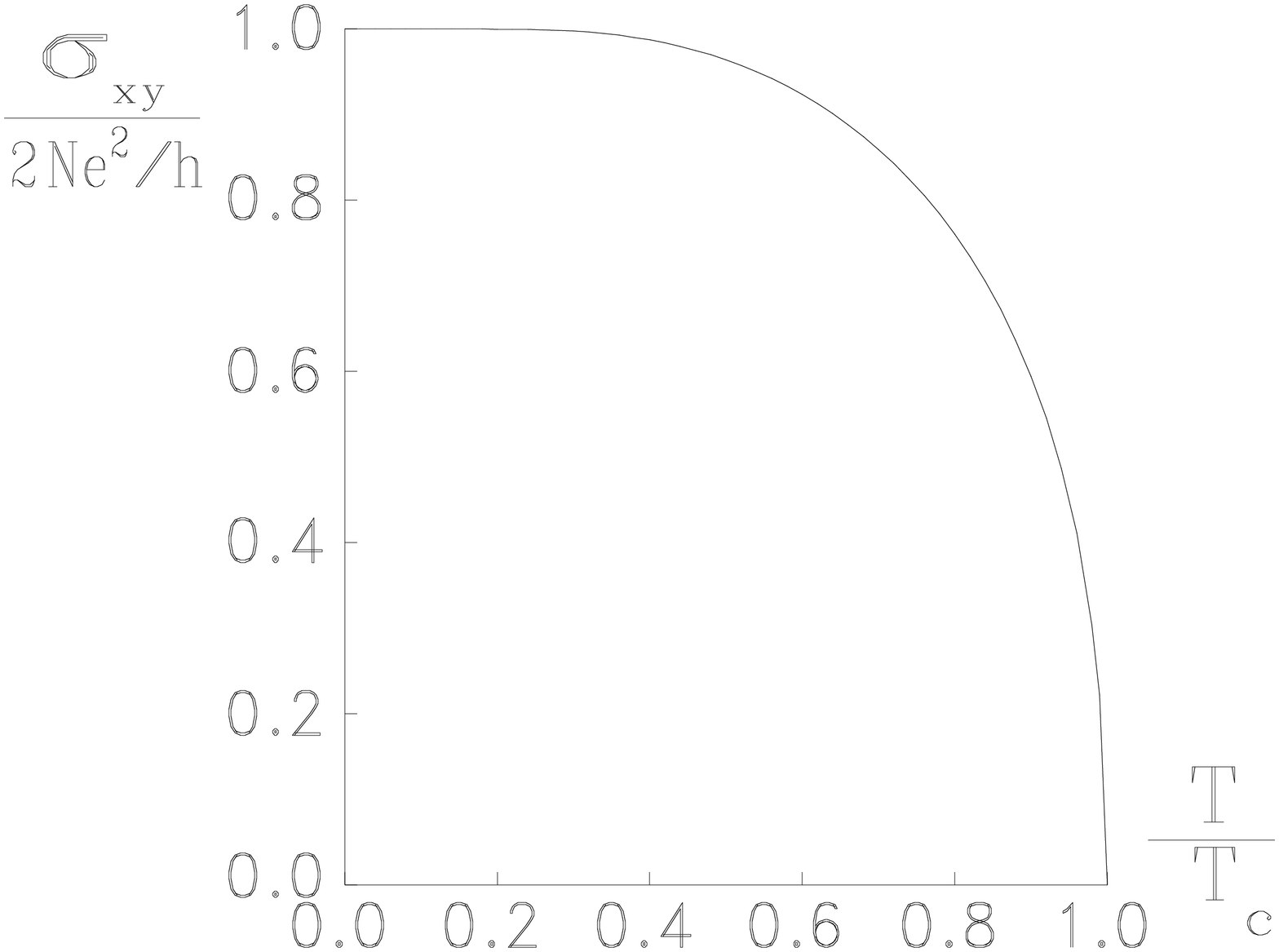}
\hspace*{\fill}
\caption{ (\textbf{a}) The temperature reduction factor $f(\Delta/T)$
of the Hall conductivity (\ref{f(T)2Ne2/h}) given by
(\ref{dynMaki}). (\textbf{b}) The Hall conductivity (\ref{f(T)2Ne2/h})
in the FISDW state as a function of temperature $T$ normalized to the
FISDW transition temperature $T_\mathrm{c}$}
\label{Fig:s_xy(T)}
\end{figure}

The function $f$ (\ref{f(T)}) depends on $\Delta/T$ and can be written
as \cite{Virosztek89,Maki90}
\begin{equation}
  f\left(\frac{\Delta}{T}\right)= \int_0^\infty \D\zeta\,
  \tanh\left(\frac{\Delta}{2T}\cosh\zeta\right) /\cosh^2\zeta \ts.
\label{dynMaki}
\end{equation}
The function $f$ is plotted in Fig.\,\ref{Fig:s_xy(T)}a.  It equals 1
at $T=0$, where Eq.\ (\ref{f(T)2Ne2/h}) reproduces the QHE, gradually
decreases with increasing $T$, and vanishes at $T\gg\Delta$.  Taking
into account that the FISDW order parameter $\Delta$ itself depends on
$T$ and vanishes at the FISDW transition temperature $T_\mathrm{c}$, it is
clear that $f(T)$ and $\sigma_{xy}(T)$ vanish at $T\rightarrow T_\mathrm{c}$,
where $\sigma_{xy}(T)\propto f(T)\propto\Delta(T)\propto\sqrt{T_\mathrm{c}-T}$.
Assuming that the temperature dependence $\Delta(T)$ is given by the
BCS theory \cite{Poilblanc86,Virosztek86}, we plot $\sigma_{xy}(T)$ in
Fig.\,\ref{Fig:s_xy(T)}b.  Strictly speaking $\sigma_{xy}(T)$ should
not vanish at $T\rightarrow T_\mathrm{c}$, but approach to the Hall
conductivity of the metallic phase.  However, as discussed in
Sect.\,\ref{Sec:Normal}, the Hall effect in the normal state is very
small, so this modification is not essential.

The function $f(T)$ (\ref{f(T)}) is qualitatively similar to the
function $f_{\rm s}(T)$ that describes temperature dependence of the
superconducting condensate density in the London case
\cite{Schrieffer-book}. Both functions equal 1 at $T=0$, but the
superconducting function behaves differently near $T_\mathrm{c}$: $f_{\rm
s}(T)\propto \Delta^2(T)\propto T_\mathrm{c}-T$. To understand the difference
between the two functions, they should be considered at small, but
finite frequency $\omega$ and wave vector $q$ \cite{Yakovenko98}.
Equations (\ref{f(T)}) and (\ref{dynMaki}) represent the dynamic limit,
where $q/\omega=0$.  This is the relevant limit in our case, because
the electric field is homogeneous in space ($q=0$), but may be
time-dependent ($\omega\neq0$).  The effective periodic potential
(\ref{1-gap}) is also time-dependent in the presence of $E_y$, as
shown in (\ref{j_x}).  On the other hand, for the Meissner effect in
superconductors, where the magnetic field is stationary ($\omega=0$),
but varies in space ($q\neq0$), the static limit $\omega/q=0$ is
relevant.  The dynamic and static limits are discussed in more detail
in \cite{Yakovenko98}.

In the derivation of (\ref{f(T)2Ne2/h}) we assumed that the wave
vector $Q_x$ (\ref{Q}) is always quantized with an integer $N$, so
that the energy gap is located at the Fermi level.  While this is the
case at $T=0$ \cite{Sengupta03}, Lebed pointed out in \cite{Lebed02}
that $Q_x$ is not necessarily quantized at $T\neq0$.  Because of the
thermally excited quasiparticles and multiple periodic potentials
present in (\ref{Fourier}), the optimal value of $N$ in (\ref{Q}) may
be non-integer, which results in deviations from the QHE.  Equation
(\ref{f(T)2Ne2/h}) was compared with the experimental temperature
dependence of the Hall effect in the FISDW state using a limited data
set in \cite{Yakovenko99} and detailed measurements in
\cite{Vuletic01}.  A good quantitative agreement was found for small
integer numbers $N\sim1$, where the QHE is well-defined.  However, for
the FISDW with bigger $N$ and lower $T_\mathrm{c}$ at lower $B$, poor
quantization of the Hall effect was found \cite{Kornilov02} at the
experimentally accessible temperatures, in qualitative agreement with
the theory of the non-quantized FISDW \cite{Lebed02,Sengupta03}.

\section{Influence of the FISDW motion on the quantum Hall effect}
\label{Sec:motion}

In the derivation of (\ref{j_x}), we assumed that the FISDW is pinned
and produces only a static periodic potential.  However, when the
FISDW is subject to a strong or time-dependent electric field, it may
move.  It is interesting to study how this motion would affect the QHE
\cite{Yakovenko96,Yakovenko98,Yakovenko93}.

Motion of the FISDW can be described by introducing a time-dependent
phase\footnote{The phase $\Theta$ in this paper has the opposite sign
to $\Theta$ in \cite{Yakovenko96,Yakovenko98}.} $\Theta$ of the FISDW
amplitude $\Delta_0$ in (\ref{H1}):
$\Delta_0\to\Delta_0\E^{-\I\Theta}$.  Then, this phase re-appears in
the off-diagonal terms in (\ref{1-gap}) and contributes to the electric
current (\ref{j_x}) along the chains
\begin{equation}
  j_x=\frac{2Ne^2}{h}\,E_y
  +\frac{2e}{2\pi b}\,\dot{\Theta}\ts,
\label{j_x+Theta}
\end{equation}
where the dot represents the time derivative.  Equation
(\ref{j_x+Theta}) needs to be supplemented with an equation of motion
for $\Theta$.  The latter can be obtained from the effective
Lagrangian of the system derived in
\cite{Yakovenko93,Yakovenko96,Yakovenko98}
\begin{equation}
  L = -\frac{Ne^2}{hc}\varepsilon_{ijk}A_iF_{jk} 
  + \frac{\hbar}{4\pi bv_{\rm F}}\dot{\Theta}^2 
  + \frac{e}{\pi b}\Theta E_x 
  + \frac{eN}{2\pi v_{\rm F}}\dot{\Theta}E_y \ts.
\label{L}
\end{equation}
Summation of over $(i,j,k)=(x,y,t)$ is implied in the first term, and
$F_{jk}$ is the electromagnetic field tensor .  The first term in
(\ref{L}) is the Chern-Simons term responsible for the QHE
\cite{Yakovenko90,Yakovenko91}.  The second term is the kinetic energy
of a moving FISDW.  The third term, well known in the theory of
density waves \cite{Gruner94}, represents potential energy in the
electric field along the chains.  The most important for us is the
last term, which describes interaction between the FISDW motion and
the electric field perpendicular to the chains
\cite{Yakovenko93,Yakovenko96,Yakovenko98}.  This term is permitted by
symmetry and has the structure of a mixed product
$\vec{v}[\vec{E}\times\vec{B}]$.  Here, $\vec{v}$ is the velocity of
the FISDW, proportional to $\dot{\Theta}$ and directed along the
chains, i.e.\ along the $x$ axis.  The magnetic field $\vec{B}$ is
directed along the $z$ axis; so, the electric field $\vec{E}$ enters
through the component $E_y$.  One should keep in mind that the
magnetic field enters the last term in (\ref{L}) implicitly, through
the integer $N$, which depends on $B$ and changes sign when $B$
changes sign.  Varying (\ref{L}) with respect to $\Theta$ and
phenomenologically adding the pinning and friction terms, we find the
FISDW equation of motion
\begin{equation}
  \ddot{\Theta}+\frac{1}{\tau}\dot{\Theta}+\omega_0^2\Theta
  =\frac{2ev_{\rm F}}{\hbar}E_x - \frac{eNb}{\hbar}\dot{E_y} \ts,
\label{fric}
\end{equation}
where $\tau$ is a relaxation time, and $\omega_0$ is the pinning
frequency.

Let us first consider the ideal case of a free FISDW without pinning
and damping.  If the electric field $E_y$ is applied perpendicular to
the chains, the last term in (\ref{fric}) induces such a motion of the
FISDW that the second term in (\ref{j_x+Theta}) exactly cancels to the
first term, and the Hall effect vanishes.  If the electric field $E_x$
is parallel to the chains, we consider the perpendicular current $j_y$
obtained by varying (\ref{L}) with respect to $A_y$:
\begin{equation}
  j_y=-\frac{2Ne^2}{h}E_x + \frac{eN}{2\pi v_{\rm F}}\ddot{\Theta}.
\label{j_y}
\end{equation}
Using (\ref{fric}) for the ideal case, we see that the two terms in
(\ref{j_y}) cancel out, and the Hall effect vanishes.  The
cancellation of the QHE by the moving FISDW is in the spirit of Lenz's
law, which says that a system responds to an external perturbation in
such a way as to minimize its effect.  Thus, the QHE exists only if
the FISDW is pinned and does not move.

\begin{figure}
\centering
\includegraphics*[width=0.4\textwidth,angle=-90]{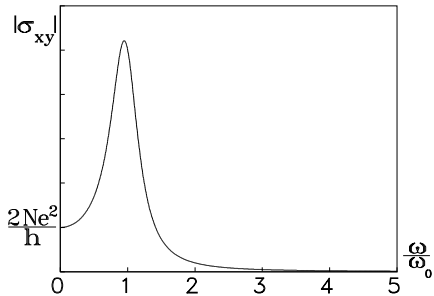}
\caption{The absolute value of the Hall conductivity $|\sigma_{xy}|$
computed from (\ref{omega}) for $\omega_0\tau=2$ as a function of
frequency $\omega$ normalized to the pinning frequency $\omega_0$}
\label{Fig:sxy(w)}
\end{figure}

In a more realistic case with pinning and damping, we solve
(\ref{fric}) by the Fourier transform from time $t$ to frequency
$\omega$ and substitute the result into (\ref{j_x+Theta}) and
(\ref{j_y}) to obtain the ac Hall
conductivity\footnote{$\sigma_{xy}(\omega)$ for a FISDW was studied in
\cite{Virosztek89}, but failed to reproduce the QHE at $\omega=0$.}
\begin{equation}
  \sigma_{xy}(\omega)=\frac{2Ne^2}{h}\frac{\omega_0^2-\I\omega/\tau}
  {\omega_0^2-\omega^2-\I\omega/\tau} \ts.
\label{omega}
\end{equation}
The absolute value $|\sigma_{xy}|$ computed from (\ref{omega}) is
plotted in Fig.\,\ref{Fig:sxy(w)} as a function of $\omega/\omega_0$
for $\omega_0\tau=2$. It is quantized at $\omega=0$ and has a
resonance at the pinning frequency.  At higher frequencies, where
pinning and damping can be neglected, and the FISDW behaves as a free
inertial system, we find that $\sigma_{xy}(\omega)\to0$.

Frequency dependence of the Hall conductivity in semiconducting QHE
systems was measured using the technique of crossed wave guides in
\cite{Kuchar,Galchenkov}.  No measurements of $\sigma_{xy}(\omega)$
have been done in (TMTSF)$_2$X thus far, but they would be very
interesting and can differentiate the QHE in the FISDW state from the
conventional QHE in semiconductors.  To give a crude estimate of the
required frequency range, we quote the pinning frequency
$\omega_0\sim$ 3 GHz $\sim$ 0.1 K $\sim$ 10 cm for a regular SDW (not
FISDW) in (TMTSF)$_2$PF$_6$ \cite{Quinlivan}.

The FISDW can be also depinned by a strong dc electric field.  In this
case, the FISDW motion is controlled by dissipation, which is
difficult to study theoretically on microscopic level.  For a steady
motion, the last terms in (\ref{fric}) and (\ref{j_y}) drop out, so we
expect no changes in $\sigma_{xy}$, but some increase in $\sigma_{xx}$
due to the sliding FISDW \cite{Virosztek89}.\footnote{In principle,
due to the presence of \vec{B}, we can phenomenologically add a term
proportional to $E_y$ to (\ref{fric}) and a term proportional to
$\dot{\Theta}$ to (\ref{j_y}).  These terms have dissipative origin
and violate the time reversal symmetry, so they cannot be obtained
from a Lagrangian.  Deriving them from the Boltzmann equation is a
difficult task.}  If $\sigma_{yy}$ also increases due to increased
dissipation and excitation of quasiparticles, then we expect that
$\rho_{xx}=\sigma_{yy}/(\sigma_{xy}^2+\sigma_{xx}\sigma_{yy})$ and
$\rho_{yy}=\sigma_{xx}/(\sigma_{xy}^2+\sigma_{xx}\sigma_{yy})$ would
increase when the FISDW starts to slide, whereas
$\rho_{xy}=\sigma_{xy}/(\sigma_{xy}^2+\sigma_{xx}\sigma_{yy})$ would
decrease.  The experimental measurements in (TMTSF)$_2$PF$_6$ are in
qualitative agreement with these expectation \cite{Balicas93},
although earlier measurements in (TMTSF)$_2$ClO$_4$ \cite{Osada87}
produced a different result.

The influence of steady motion of a regular charge-density wave (not
FISDW) on the Hall conductivity was studied theoretically in
\cite{Artemenko84b} and experimentally in
\cite{Ong81,Artemenko84a,Forro86}.  In this case, there is no QHE
contribution from the condensate, and the effect is primary determined
by the thermally excited normal carriers.

\section{Chiral edge states}
\label{Sec:edges}

\begin{figure}
\centering
\includegraphics*[width=0.5\linewidth]{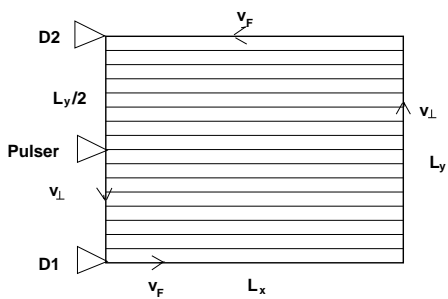}
\caption{Chiral edge modes circulate around the sample with the
velocities $v_{\perp}$ and $v_\mathrm{F} $ in the direction of the
arrows.  The thin parallel lines represent conducting chains of a Q1D
material.  In the proposed time-of-flight experiment, a pulse from the
pulser is detected at different times $t$ and $t'$ by the detectors D1
and D2}
\label{Fig:chiral}
\end{figure}

Thus far, we studied the QHE in the bulk.  Generally, a system with
the integer QHE characterized by the number $N$ is expected to have
$N$ chiral edge states \cite{Halperin82,Kane97,Hatsugai93}.  These
electron states are localized at the boundaries of the sample and
circulate along the boundary with some velocity $v$, as shown in
Fig.\,\ref{Fig:chiral}.  Excitations with the opposite sense of
circulation are absent, so these states is chiral.  Theory of the edge
states in multilayered QHE systems was discussed phenomenologically in
\cite{Chalker95,Balents96}.

The QHE can be equivalently formulated in terms of the chiral edge
states \cite{Halperin82,Kane97}.  Suppose a small electric voltage
$V_y$ is applied across the sample.  It produces a difference of
chemical potentials between the opposite edges of the sample.  The
electron states in the bulk of the sample are gapped, so they would
not respond to this perturbation.  However, the edge modes are
gapless, so the difference of chemical potentials produces imbalance
$\udelta n=2eV_y/hv$ between the occupation numbers of the chiral
modes at the opposite edges.  Here we utilized the 1D density of
states $2/hv$, accounting for two spin projections.  The chiral modes
at the opposite edges propagate in the opposite directions, so the
population imbalance between them generates the net edge current $I_x$
in the $x$ direction:
\begin{equation}
  I_x=evN\,\udelta n=evN\,\frac{2eV_y}{hv}=\frac{2Ne^2}{h}\,V_y \ts.
\label{Ix}
\end{equation}
Equation (\ref{Ix}) represents the QHE, this time for the Hall
conductance, rather than conductivity (\ref{s_xy}), which are the same
in 2D.  Notice that the velocity $v$ of the chiral edge states cancels
out in (\ref{Ix}), and only their number $N$ enters the final formula.
The bulk and edge formulations of the QHE are equivalent.  Whether the
Hall current actually flows in the bulk or along the edge depends on
where the voltage drops in the sample, but the overall Hall
conductance does not depend on this.

Below we discuss the structure of the edge states for the FISDW
\cite{Yakovenko96,Sengupta01}.\footnote{The edge states in the normal
phase of Q1D conductors in a magnetic field were studied in
\cite{Goan98}.}

\subsection{Edges perpendicular to the chains}
\label{Sec:perpendicular}

First, let us consider a 1D density wave occupying the positive
semi-space $x>0$ with an edge at $x=0$.  In this case, the wave
function (\ref{+-k_F}) must vanish at the edge: $\psi(x=0)=0$, so
$\psi_+(x=0)=-\psi_-(x=0)$.  With this boundary condition, the
Hamiltonian (\ref{1-gap}) also admits a localized electron state, in
addition to (\ref{E(p_x)}), with the energy $|E|<\Delta$ inside the
gap:
\begin{equation}
  E = -\Delta \cos\varphi \ts, \quad
  \psi_{\pm} = \pm \frac{\E^{-\kappa x}}{\sqrt\kappa} \ts, \quad 
  \kappa = -\frac{\sin\varphi}{\xi} \ts, \quad
  \xi=\frac{\hbar v_\mathrm{F}
}{\Delta} \ts.
\label{onedr}
\end{equation}
The wave function (\ref{onedr}) exponentially decays into the bulk at
a length of the order of the coherence length $\xi$.  Equation
(\ref{onedr}) is meaningful only when $\kappa>0$, so the localized
state exists on the left edge only for $\pi<\varphi<2\pi$.  However, a
solution with $\kappa<0$ is appropriate for the right edge at the
opposite end of the sample.  The edge state (\ref{onedr}) is
mathematically similar to the localized state at a kink soliton in a
density wave \cite{Brazovskii}.

\begin{figure}
\centering
\includegraphics*[width=0.55\linewidth]{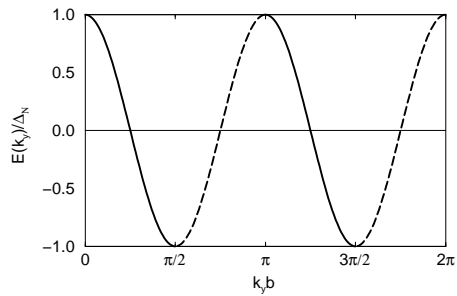}
\caption{Energy dispersion $E/(-\Delta)$ (\ref{twodr}) of the electron
states localized at the right (solid lines) and left (dashed lines)
edges of the sample as a function of the transverse momentum $k_y$ for
$N=2$}
\label{Fig:sin}
\end{figure}

Now let us consider a FISDW occupying the positive semi-space $x>0$
along the chains and extended in the $y$-direction.  In this case, the
phase $\varphi=Nbk_y$ in (\ref{1-gap}) depends on $k_y$.  Substituting
$\varphi(k_y)$ into (\ref{onedr}), we find \cite{Sengupta01}
\begin{equation}
  E(k_y) = -\Delta \cos(Nk_yb) \ts, \quad 
  \kappa = -\frac{\sin(Nk_yb)}{\xi} \ts, \quad 
  \psi_{\pm} = \pm \frac{\E^{\I k_yy -\kappa x}}{\sqrt\kappa} \ts.
\label{twodr}
\end{equation}
The single bound state (\ref{onedr}) is now replaced by the band
(\ref{twodr}) of the edge states labeled by the wave vector $k_y$
perpendicular to the chains.  These states (\ref{twodr}) are localized
along the chains and extended perpendicular to the chains.  At the
left edge of the sample, we require that $\kappa\propto-\sin(Nk_yb)$
is positive, which gives $N$ branches of the edge states in the
transverse Brillouin zone $0<k_yb<2\pi$.  The complementary $N$
branches of the edge states, determined by the condition $\kappa<0$,
exist at the right edge.  The dashed and solid lines in
Fig.\,\ref{Fig:sin} show the energy dispersion $E(k_y)$ (\ref{twodr})
of the states localized at the left and right edges for $N=2$.  It is
clear that the group velocities of the edge states $\partial
E(k_y)/\hbar\partial k_y$ have opposite signs for the left and right
edges.  Thus, they carry a surface current around the sample, as
indicated by the arrows in Fig.\,\ref{Fig:chiral}.  The sense of
circulation is determined by the sign of $N$, which is controlled by
the sign of the magnetic field $B$.

The edge states bands are filled up to the Fermi level in the middle
of the gap, and their group velocity at the Fermi level is
\begin{equation}
  v_{\perp} = \frac{1}{\hbar} \left.
  \frac{\partial E(k_y)}{\partial k_y}\right|_{E=0} 
  = \frac{Nb\Delta}{\hbar} \ts, 
\label{v_perp}
\end{equation}
The velocity $v_{\perp}$ is quite low, because it is proportional to
the small FISDW gap $\Delta$, so $v_{\perp}\ll v_\mathrm{F}
$.

\subsection{Edges parallel to the chains}
\label{Sec:parallel}

Now let us discuss the edges parallel to the chains.  The effective
Hamiltonian (\ref{1-gap}) operates on the electron wave functions
$\psi_{\pm}'(k_x,k_y)$ labeled by the momentum $k_y$.  The bulk energy
spectrum (\ref{E(p_x)}) of (\ref{1-gap}) is degenerate in $k_y$.
Thus, we can use the Wannier wave functions
$\psi_{\pm}'(k_x,M)=\int\E^{\I k_yMb}\psi_{\pm}'(k_x,k_y)\,\D
k_y/2\pi$ as a new basis \cite{Yakovenko87b,Goan98}.  The wave
function $\psi_{\pm}'(k_x,M)$ is localized across the chains around
the chain with the number $M$.  Introducing the destruction operators
$\hat{a}_\pm(p_x,M)$ for this basis, we can rewrite the Hamiltonian
(\ref{1-gap}) in the following form \cite{Yakovenko87b}
\begin{eqnarray}
  &&\hat{H}'=\int\frac{\D k_x}{2\pi}\sum_M\,
  v_\mathrm{F}
p_x\,[\hat{a}^+_+(p_x,M)\,\hat{a}_+(p_x,M)
  -\hat{a}^+_-(p_x,M)\,\hat{a}_-(p_x,M)] \nonumber 
\\ &&
  {}+\Delta\,[\hat{a}^+_+(p_x,M+N)\,\hat{a}_-(p_x,M)
  +\hat{a}^+_-(p_x,M)\,\hat{a}_+(p_x,M+N)] \ts.
\label{MN}
\end{eqnarray}
As a consequence of the $k_y$-dependent phase in the off-diagonal
terms in (\ref{1-gap}), the Hamiltonian (\ref{MN}) represents pairing
between the $+k_\mathrm{F}$ and $-k_\mathrm{F}$ states localized at
the different chains $M+N$ and $M$.  For the chains in the bulk of the
crystal, this pairing results in the gapped energy spectrum
(\ref{E(p_x)}).  However, the states at the edges are exceptional.
The $+k_\mathrm{F}$ states on the first $N$ chains on one side of the
crystal and the $-k_\mathrm{F}$ states on the last $N$ chains on the
other side of the crystal do not have partners to couple with, so
these states remain ungapped \cite{Yakovenko96}.  Thus, one side of
the sample possesses $N$ gapless chiral modes propagating along the
edge with the velocity $+v_\mathrm{F} $, and the other side has $N$
gapless chiral modes propagating in the opposite direction with the
velocity $-v_ F$, as shown in Fig.\,\ref{Fig:chiral}.

\subsection{Possibilities for experimental observation of the 
  chiral edges states}
\label{Sec:edge-experimental}

Let us estimate parameters of the chiral edge states.  The activation
energy in the FISDW state with $N=1$ was found to be $2\Delta=6$~K in
$\rm(TMTSF)_2ClO_4$ at $B=25$ T \cite{Chamberlin}.  Substituting
$\Delta=3$~K and $N=1$ into (\ref{v_perp}), we find
$v_{\perp}=300$~m/s, which is three orders of magnitude lower than
$v_\mathrm{F} =10^5$~m/s \cite{Takahashi05}.  Despite the big
difference in velocities, the total currents of the parallel and
perpendicular edge states are the same.  Indeed, the slow
perpendicular states with $v_{\perp}=Nb\Delta/\hbar$ have the large
width $\xi=\hbar v_\mathrm{F} /\Delta$, whereas the fast parallel
states with the velocity $v_\mathrm{F} $ have the narrow width $Nb$.
The total edge current $I$ carried by the perpendicular states
(\ref{twodr}) is
\begin{equation}
  I = \frac{2Ne}{h}\int_0^{\pi/2bN}\frac{\partial E(k_y)}{\partial
  k_y}\,\D k_y =\frac{2Ne\Delta}{h}=\frac{ev_{\perp}}{\pi b} =
  20\:{\rm nA} \ts.
\label{I_perp}
\end{equation}
The same current is carried along the chains by the difference between
the gapped and ungapped branches of the electron dispersion
(\ref{E(p_x)})
\begin{equation}
  I = \frac{2Ne}{h}\int_{-\infty}^0\left(\frac{\partial
  E(p_x)}{\partial p_x}-v_\mathrm{F} \right)\D p_x
  =\frac{2Ne\Delta}{h} \ts.
\label{I_||}
\end{equation}
It is tempting to use (\ref{I_perp}) to calculate magnetization of the
sample.  However, there are additional contributions to the total
magnetization coming from the edge states inside the energy gaps
opened by the neglected terms in (\ref{Fourier}) below the Fermi
level.  Magnetization of the FISDW state was calculated in
\cite{Montambaux88} using the bulk free energy and was measured in
$\rm(TMTSF)_2ClO_4$ in \cite{Naughton85}.

The most convincing demonstration of the edge states would be the
time-of-flight experiment \cite{Sengupta01} analogous to that
performed in GaAs in \cite{Ashoori92,Ernst97}.  The experimental setup
is sketched in Fig.\,\ref{Fig:chiral}.  An electric pulse is applied
by the pulser at the center of the edge perpendicular to the chains.
The pulse travels counterclockwise around the sample.  For the typical
sample dimensions $L_x=2$ mm and $L_y=0.2$ mm, we find the times of
flight $t_x=L_x/v_\mathrm{F} =8$~ns and
$t_y=L_y/v_{\perp}=0.67$~$\umu$s using the values of $v_{\perp}$ and
$v_\mathrm{F} $ quoted above for $N=1$.  Thus, the pulse will reach
the detector D1 at the time $t=L_y/2v_{\perp}$ and the other detector
D2 at the longer time $t'=3t + 2L_x/v_\mathrm{F} \approx3t$.  The
difference between the arrival times $t$ and $t'$ is a signature of
the chiral edge states.  The flight time $t$ should exhibit
discontinuities at the FISDW phase boundaries due to discontinuity of
both $N$ and $\Delta$ affecting $v_{\perp}$ in (\ref{v_perp}).  The
pulse must be shorter than $t=0.33$ $\umu$s for clear resolution and
longer than $\hbar/\Delta=2.6$ ps, so that only the low-energy
excitations are probed.

Existence of the edge states was confirmed in a multilayered GaAs
system by showing that the conductance $G_{zz}$ perpendicular to the
layers is proportional to the number of the edge states and the
perimeter of the sample \cite{Druist}.  A similar experiment in
$\rm(TMTSF)_2AsF_6$ produced inconclusive results \cite{Uji}.

The specific heat per layer $C_\E$ of the gapless edge excitations is
proportional to temperature at $T\ll\Delta$ \cite{Sengupta01}:
\begin{equation}
  \frac{C_\E}{T} = \frac{N\pi}{3\,\hbar} \left(\frac{2 L_y}{v_{\perp}}
  + \frac{2 L_x}{v_\mathrm{F}
} \right)
  \approx\frac{2\pi}{3\Delta}\,\frac{L_y}{b} \approx 3 \times
  10^{-3}\:\frac{\rm mJ}{\rm K^2\,mole} \ts,
\label{C/T}
\end{equation}
where the dominant contribution comes from the edges perpendicular to
the chains.  The ratio of $C_\E/T$ (\ref{C/T}) to the bulk specific
heat in the normal state $C_\mathrm{b}^{\rm(n)}/T=\pi L_xL_y/3\hbar
bv_\mathrm{F} $ is roughly equal to the ratio of the volumes occupied
by the bulk and edge states:
$C_\mathrm{b}^{\rm(n)}/C_\E=L_x/2\xi\approx10^3$.  The experimentally
measured $C_\mathrm{b}^{\rm(n)}/T$ is 5~mJ/(K$^2$\,mole)
\cite{Scheven95}.  In the FISDW state, the bulk specific heat is
exponentially suppressed because of the energy gap $\Delta$ and
becomes smaller than $C_\E$ (\ref{C/T}) at sufficiently low
temperatures $T\le\Delta/14$ \cite{Sengupta01}.  This regime could
have been possibly achieved in the specific heat measurements
\cite{Scheven95} performed at $T=0.32$ K and $B=9$ T in
$\rm(TMTSF)_2ClO_4$.  According to (\ref{C/T}), $C_\E/T$ must be
discontinuous at the boundaries between the FISDW phases, where
$\Delta$ changes discontinuously \cite{Montambaux88}.  Notice that $N$
cancels out in (\ref{C/T}), in contrast to the phenomenological model
\cite{Balents96}.

Other possibilities for experimental observation of the edge states
are discussed in \cite{Kwon03}.  We would like to mention that
non-chiral midgap edge states are expected to exist in the
superconducting $p$-wave state of $\rm(TMTSF)_2X$ \cite{midgap}.

\section{Generalization to the three-dimensional quantum Hall effect}
\label{Sec:3D}

Experiments show that the FISDW state in $\rm(TMTSF)_2X$ depends only
on the $B_z$ component of a magnetic field.  Thus, although the
$\rm(TMTSF)_2X$ crystals are three-dimensional (3D), they can be
treated as a collection of essentially independent 2D layers, and the
QHE can be studied within a 2D theory. Nevertheless, let us discuss a
generalization of this theory to the 3D case.

Suppose a magnetic field has the $B_z$ and $B_y$ components
perpendicular to the chains.  As Lebed pointed out in \cite{Lebed},
inserting them into (\ref{dispersion}) via the Peierls--Onsager
substitution creates two periodic potentials in the $x$ direction,
$2t_b\cos(k_yb-G_1x)$ and $2t_c\cos(k_zd+G_2x)$, with the wave vectors
$G_1$ and $G_2$.  Thus, a FISDW may form with the wave vector
\cite{Montambaux89}
\begin{equation}
  \quad Q_x=2k_\mathrm{F} - N_1G_1 - N_2G_2 \ts, \quad 
  G_1=ebB_z/\hbar c \ts, \quad G_2=edB_y/\hbar c \ts.
\label{Q2}
\end{equation}
Repeating the derivation of Sect.\,\ref{Sec:FISDW} and introducing the
electric field components $E_x$ and $E_y$ as in (\ref{j_x}), we find
the electric current per one chain \cite{Sun94,Hasegawa95}
\begin{equation}
  I_x=\frac{2e}{2\pi}\,\frac{\partial\varphi}{\partial t}
  =\frac{2e^2}{h}(N_1bE_y - N_2dE_z) \ts.
\label{I_x}
\end{equation}
Using the current density per unit area $j_x=I_x/bd$, we rewrite
(\ref{I_x}) as
\begin{equation}
  \vec{j} = \frac{2e^2}{h}\vec{E}\times\vec{K}\ts, \quad
  \vec{K}=\left(0,\frac{N_2}{b},\frac{N_1}{d}\right) \ts.
\label{j_x2}
\end{equation}
Equation (\ref{j_x2}) with an integer vector \vec{K} belonging to the
reciprocal crystal lattice represents the general form of the QHE in a
3D system \cite{Halperin87}.  This formula was applied to the 3D FISDW
(\ref{Q2}) in \cite{Montambaux89,Hasegawa95,Koshino02} and to general
lattices in \cite{Montambaux90,Kohmoto92,Koshino01}.  The edge states
picture, presented in Sect.\,\ref{Sec:edges}, was generalized to the
3D FISDW in \cite{Halperin02}.

Although the 3D FISDW (\ref{Q2}) does not realize in $\rm(TMTSF)_2X$,
it may occur in other families of Q1D materials.  It would be very
interesting to investigate whether FISDW exists in the Q1D material
(DI--DCNQI)$_2$Ag, where $t_b=t_c$ by crystal symmetry, so there is no
layered structure \cite{Itou04}.

\section{Conclusions and open questions}
\label{Sec:Conclusions}

In summary, the QHE in the FISDW state is a direct consequence of the
quantization and magnetic field dependence of the wave vector $Q_x$
(\ref{Q}).  As a result, $N$ completely filled Landau bands are
maintained between $Q_x$ and $2k_\mathrm{F}$, whereas the Fermi sea
plays the role of a reservoir.  More rigorously, the QHE can be
obtained as a topological invariant in terms of the winding number of
a phase in the Brillouin zone.  The topological approach is
particularly useful when several FISDWs coexist.  The Hall effect can
be also viewed as the Fr\"ohlich current of an effective density
wave.\footnote{The theory presented in Sect.\,\ref{Sec:FISDW} and
(\ref{j_x}) was also applied to quantized adiabatic transport induced
by surface acoustic waves in carbon nanotubes \cite{Talyanskii01}.}
This allows us to derive its temperature dependence within a two-fluid
picture.  Motion of the FISDW cancels the Hall effect in the ideal
case of a free FISDW and at high frequency.  The QHE can be also
formulated in terms of the chiral edge states circulating around the
sample with low speed across the chains and high speed along the
chains.  The QHE can be also generalized for a three-dimensional
FISDW.  While the temperature dependence of the QHE was measured
experimentally and found to be in agreement with the theory, the other
theoretical results, such as the frequency dependence of $\sigma_{xy}$
and the existence of the edge states, await experimental verification.
It would be interesting to search for a 3D FISDW in (DI--DCNQI)$_2$Ag.

Below we list some open questions in the theory of the QHE and the
FISDW.  One problems is that, in experiment, the dissipative
components $\sigma_{xx}$ and $\sigma_{yy}$ tend to saturate at small,
but finite values in the limit $T\to0$ \cite{Jerome88,Valfells96}.
The reason for this is not completely clear, but may be due to
impurity scattering \cite{Azbel87}.  As a result, the Hall effect
quantization is not as good as in semiconducting systems, especially
at lower magnetic fields.

Predicting the sequence of $N$ as a function of $B$ is a task for the
FISDW theory.  For a simple model, the theory gives consecutive
numbers \cite{Montambaux85,Montambaux88}, as observed experimentally
at higher pressures in $\rm(TMTSF)_2PF_6$
\cite{Chaikin92,Hannahs88,Jerome88,Valfells96}.  However, at lower
pressures, a complicated sequence of $N$ with multiple sign reversals
and a bifurcation in the $B$--$T$ phase diagram is observed
\cite{Balicas95,Cho99}.  This sequence is not fully understood,
although theoretical attempts have been made
\cite{Lebed90,Yakovenko94a,Dupuis98b,Montambaux96,Dupuis98a}.

Developing a detailed theory of the FISDW in $\rm(TMTSF)_2ClO_4$ and
$\rm(TMTSF)_2ReO_4$ is even more difficult because of the period
doubling in the \vec{b} direction due to the anion ordering in these
materials.  Experiments show a complicated phase diagram for
$\rm(TMTSF)_2ClO_4$ \cite{McKernan95,Scheven95}.  Numerous theoretical
scenarios for the FISDW in $\rm(TMTSF)_2ClO_4$ are reviewed by
S.~Haddad et al. in this book and in \cite{Haddad05}.  Particularly
puzzling is the behavior of $\rm(TMTSF)_2ClO_4$ in very strong
magnetic fields between 26 and 45 T, where it is supposed to be in the
FISDW phase with $N=0$, i.e.\ with zero Hall effect.  Instead, the
Hall coefficient and other quantities show giant oscillations as a
function of the magnetic field \cite{Uji05}.  This phenomenon is not
fully understood, but may be related to the soliton theory
\cite{Lebed97}.



\printindex
\end{document}